\newcommand{\cc}{{~\rm cm^{-3}}}

\newcommand{\s}{{~\rm s}}
\newcommand{\km}{{~\rm km}}
\newcommand{\g}{{~\rm g}}
\newcommand{\K}{{~\rm K}}
\newcommand{\erg}{{~\rm erg}}
\newcommand{\yr}{{~\rm yr}}
\newcommand{\Myr}{{~\rm Myr}}
\newcommand{\pc}{{~\rm pc}}
\newcommand{\kpc}{{~\rm kpc}}

\documentclass[12pt,preprint]{aastex}
\shortauthors{Sternberg \& Soker}

\begin{document}

\title{RISING JET-INFLATED BUBBLES IN CLUSTERS OF GALAXIES}

\author{Assaf Sternberg\altaffilmark{1},  Noam Soker\altaffilmark{1}}

\altaffiltext{1}{Department of Physics,
Technion$-$Israel Institute of Technology, Haifa 32000, Israel;
phassaf@techunix.technion.ac.il; soker@physics.technion.ac.il}

\begin{abstract}
We conduct {{{{two-dimensional axisymmetric (referred to as 2.5D)}}}} 
hydrodynamical numerical simulations of bubble
evolution in clusters of galaxies. We inflate bubbles using slow, massive
jets with a wide opening angle, and follow their evolution as they rise
through the intra-cluster medium (ICM). We find that these jet-inflated
bubbles are quite stable, and can reach large distances in the cluster while
still maintaining their basic structure. The stability of the jet-inflated
bubble comes mainly from the dense shell that forms around it during it's
inflation stage, and from the outward momentum of the bubble and the shell.
On the contrary, bubbles that are inserted by hand onto the grid and not
inflated by a jet, i.e., an \emph{artificial bubble}, lack these stabilizing
factors, therefore, they are rapidly destroyed. The stability of the
jet-inflated bubble removes the demand for stabilizing magnetic fields in
the bubble.
\end{abstract}

{{\it Subject headings:} (galaxies:) cooling flows galaxies: clusters:
 general galaxies: jets}

\section{INTRODUCTION}
\label{sec:intro}

Many clusters of galaxies harbor bubbles (cavities) devoid of X-ray emission,
e.g., Perseus (Fabian et al. 2000) and Abell 2052 (Blanton et al. 2001).
These low density bubbles are inflated by the jets launched by the active
galactic nuclei (AGN) sitting at the centers of these clusters.

The estimated ages of most bubbles (Birzan et al. 2004) is about one order of
magnitude larger than the characteristic time of the Rayleigh-Taylor (RT)
instability to destroy them. This observation has prompted many authors (e.g.,
Br\"uggen \& Kaiser 2001; Kaiser et al. 2005; Jones \& De Young 2005) to
invoke an ordered magnetic field at the edge of the bubble,
{{{{or to consider the effects of viscosity (Reynolds et al. 2005),}}}} to
stabilize the bubble against the RT instability.
Indeed, numerical simulations of non-magnetic bubble evolution show them to
be disrupted quite rapidly (e.g., Br\"uggen 2003; Br\"uggen \& Kaiser 2001;
Jones \& De Young 2005; Pavlovski et al. 2008;
Reynolds et al. 2005; Robinson et al. 2004; {{{{Ruszkowski et al. 2004a, 
2004b,}}}} 2008). Adding magnetic field make
the bubble more stable (e.g., Jones \& De Young 2005; Robinson, K. et al.
2004; {{{ Ruszkowski et al. 2007). }}} However, it is not clear if magnetic
fields can indeed supply the required stability (Ruszkowski et al. 2007).
In the simulations cited above the bubble where injected at off-center
locations by a prescribed numerical procedure. This is of course
not the way bubbles are formed in clusters. {{{ Evidence suggests that
these bubble are formed by jets. }}} We term these type of bubbles that were
inserted numerically \emph{artificial bubbles}. Different in that respect
are the tower jet model simulations conducted by Nakamura et al. (2006; also
Diehl et al. 2008), which did follow the evolution of a jet and a bubble.
Nonetheless, we find some problems with this model, e.g., the angular
momentum that is assumed in the jet is too large.

Jets can inflate bubbles if their opening angle is large (i.e., wide jets),
or if they are narrow but their axis changes its direction (Soker 2004, 2006;
Sternberg et al. 2007; Sternberg \& Soker 2008). The change in direction can
result from precession (Soker 2004, 2006; Sternberg \& Soker 2008), random
change (Heinz et al. 2006), or a relative motion  between the ICM and the
AGN (Loken et al. 1995; Soker \& Bisker 2006;
Rodr\'iguez-Mart\'inez et al. 2006). The outcome of a wide jet and a rapidly
precessing jet is the same (Sternberg \& Soker 2008). For numerical
reasons we conduct a study where we inflate bubbles with wide jets, and not
with precessing jets.
{{{{We use jets with Mach number of 10. Much faster jets form elongated 
bubbles, rather than `fat' bubbles, i.e., bubbles with axes ratio close to 
unity. In addition, much faster jets cause back flow of hot gas that fills 
the region between the two bubbles, and therefore lead to the formation of 
one elongated bubble (Sternberg et al. 2007).}}}}
In this \emph{Letter} we limit our-self to comparing the evolution of
\emph{jet-inflated bubbles} (i.e., bubbles inflated solely by jets and not by
a prescribed numerical procedure) with artificial bubbles, in the goal of
emphasizing the necessity to inflate bubble self-consistently. For that, the
description of the numerical code is brief, and we present only one case out
of the several we simulated.

\section{NUMERICAL METHOD AND SETUP}
\label{sec:numerics}

The simulations were performed using the \emph{Virginia Hydrodynamics-I}
code (VH-1; Blondin et al. 1990; Stevens et al. 1992), as described in
Sternberg et al. (2007). Radiative cooling was not included.
We study a three-dimensional axisymmetric flow with a 2D grid {{{{(referred 
to as 2.5D)}}}}. We simulate half of the meridional plane 
using the two-dimensional version of the code in spherical coordinates.
The symmetry axis of all plots shown in this paper is along the x (horizontal)
axis. Due to the fact that this is a \emph{Letter}, in order to minimize
the space required to show the plots, and due to the fact that the results
of both halves of the simulated domain were either identical or exhibit the
same large scale and stability behavior, we exhibit only a quarter of the
meridional plane (half of the simulated domain).

In contrast to our previous papers, where gravity was omitted, we added
gravity to the simulations. We assume a dark matter halo with a density
profile as is given in Navarro et al. (1996). The dark matter is
not affected by the evolution of the baryonic matter, therefore, the
gravitational potential is constant,
\begin{equation}
\Phi_{NFW}(r) = 4\pi r_s^2\delta_c\rho_{crit}G\left[1-
\frac{\ln(r/r_s+1)}{r/r_s}\right],
\label{grav}
\end{equation}
where $r_s=\frac{r_v}{c}$ is a scale radius, $r_v$ is the virial radius, $c$
is the concentration factor, $\delta_c=\frac{200}{3}\frac{c^3}{\ln(1+c)-
\frac{c}{1+c}}$, and $\rho_{crit}=\frac{3H^2}{8\pi G}$ is the critical density
of the universe at $z=0$.

We set the {{{{initial}}}} density profile of the gas to maintain
hydrostatic equilibrium
(e.g., Makino et al. 1998), i.e.,
\begin{equation}
\rho_{gas}(r)=\rho_{gas,0}e^{-b}\left(\frac{r}{r_s}+1\right)^\frac{b}{r/r_s},
\end{equation}
where $b\equiv 4\pi r_s^2\delta_c\rho_{crit}\mu m_pG/kT_v$.
$T_v=\frac{\gamma G\mu m_pM_v}{3k_Br_v}$ is the virial temperature, and
$M_v$ is the virial mass.
The unperturbed ICM temperature is $2.7 \times 10^7 \K$.
We use a $256\times256$ grid, evenly spaced in the azimuthal direction. In
the radial direction, the grid was partitioned using a geometric series with
a common ratio of 1.0015 allowing for better resolution in the inner part of
the simulated domain.
More detail will be given in a forthcoming paper where an extended parameter
space will be explored.

We simulate the evolution of a bubble inflated by a jet, and
compare it to the evolution of bubble introduced manually to the grid
(an artificial bubble).
The jet is injected at a radius of $0.1 \kpc$,
with constant mass flux of $\dot M_j= 5 M_\odot \yr^{-1}$ (per one jet)
and a constant radial velocity of $v_j= 7750 \km \s^{-1}$,
inside a half opening angle of $\alpha = 70^\circ$.
The total power of the two jets is $\dot E_{2j}= 2\times10^{44} \erg \s^{-1}$.
The jet was active for a period of $\Delta t_j= 10 \Myr$,
from $t=-10 \Myr$ until $t=0$, when the jet was completely shut-off.

The spherical artificial bubbles was inserted in its full size at $t=0$.
with its volume and (constant) density about equal to that of the jet-inflated
bubble at $t=0$. The initial bubble's radius and density were set to
$R_{ab0}= 4.5\pc$, and $\rho_{ab0}=10^{-26} \g \cc$, respectively. We made
several tests. We ran cases with initial artificial-bubble density three
times larger, and three times smaller, than $\rho_{ab0}$. We also ran cases
with numerical grid with half and twice the resolution (i.e., $128\times128$
and $512\times512$ respectively) compared to our standard resolution. We
were witness to only small differences between the runs with different
resolutions and initial bubble density. Namely, our results and conclusions
are not sensitive to the numerical details.

{{{ With the parameters used here the bubble temperature is $\sim 10$ times 
higher than the ambient temperature. We can use a somewhat higher jet 
velocity that will result in a lower bubble density (Hinton et al. 2007) 
and in a higher bubble temperature (see Sternberg et al. 2007 and Sternberg 
\& Soker 2007). This will be examined in a future paper. }}}

\section{RESULTS}
\label{sec:results}
\begin{figure}  
\hskip -2.8 cm  
\vskip 4.0 cm  
{\includegraphics[scale=0.5]{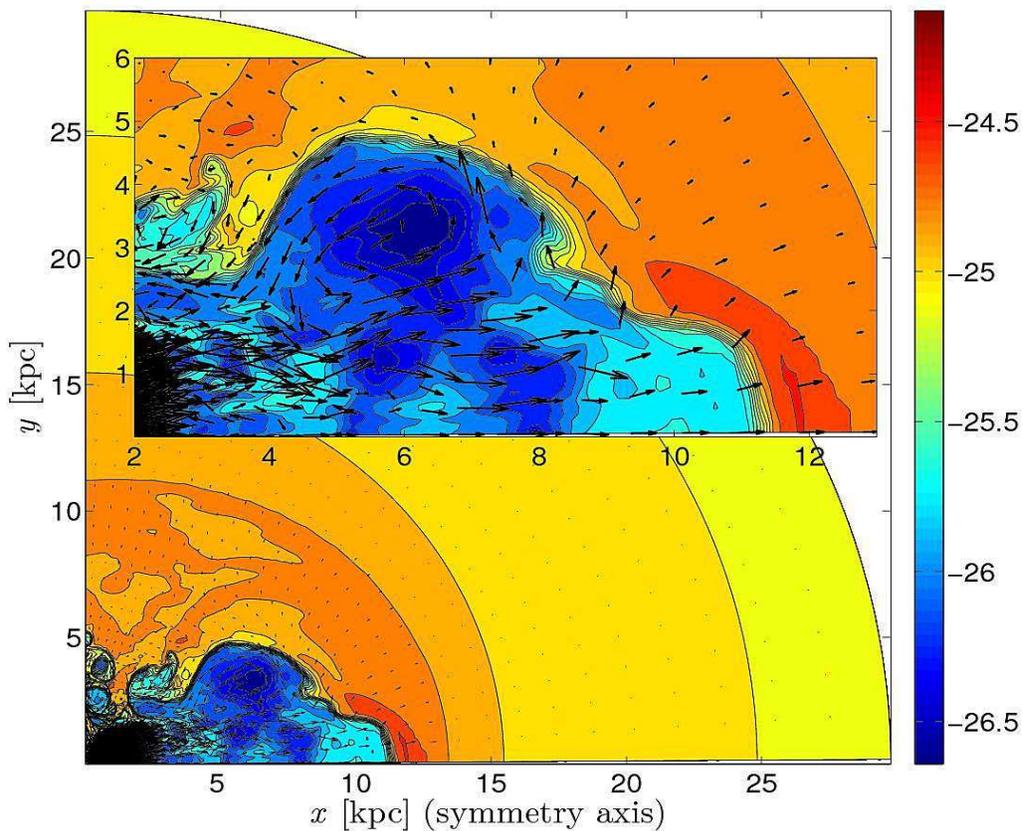}}  
\caption{Density and velocity map for the jet-inflated bubble at $t=0$,
the time the jet is shut off. The density scale is on the right in
logarithmic scale and cgs units. The arrows represent the velocity of the
flow:
$0.1c_s<v_j \leq 0.5c_s$ (shortest), $0.5c_s<v_j \leq c_s$,
$c_s<v_j \leq 5c_s$, and $5c_s<v_j \leq 10c_s$ (longest in this case).
Here $c_s=775 \km \s^{-1}$ is the sound speed of the undisturbed ICM.
 }
\label{fig:jeti}
\end{figure}

In Figure \ref{fig:jeti} we show the density and velocity map of the
jet-inflated bubble at $t=0$. There are some flow structures which
characterize the jet-inflated bubble, and that are not present in the
case of the artificial bubble. As we shall see below, these
are crucial for the future evolution of the bubble.
(1) There is a dense shell around the bubble.
(2) The bubble and the dense shell have radial momentum. In particular,
the shell's front is relatively dense and has an outward velocity.
(3) There is a circular flow, a vortex, around the lowest density region
in the bubble.
(4) Although the jet was shutdown, the jet material along the path form the
center to the bubble still exists. In reality, the jet is gradually shut down
(even if on a short time scale), which result in low density material behind
the bubble, hence, this is a real feature.

\begin{figure}[!ht]
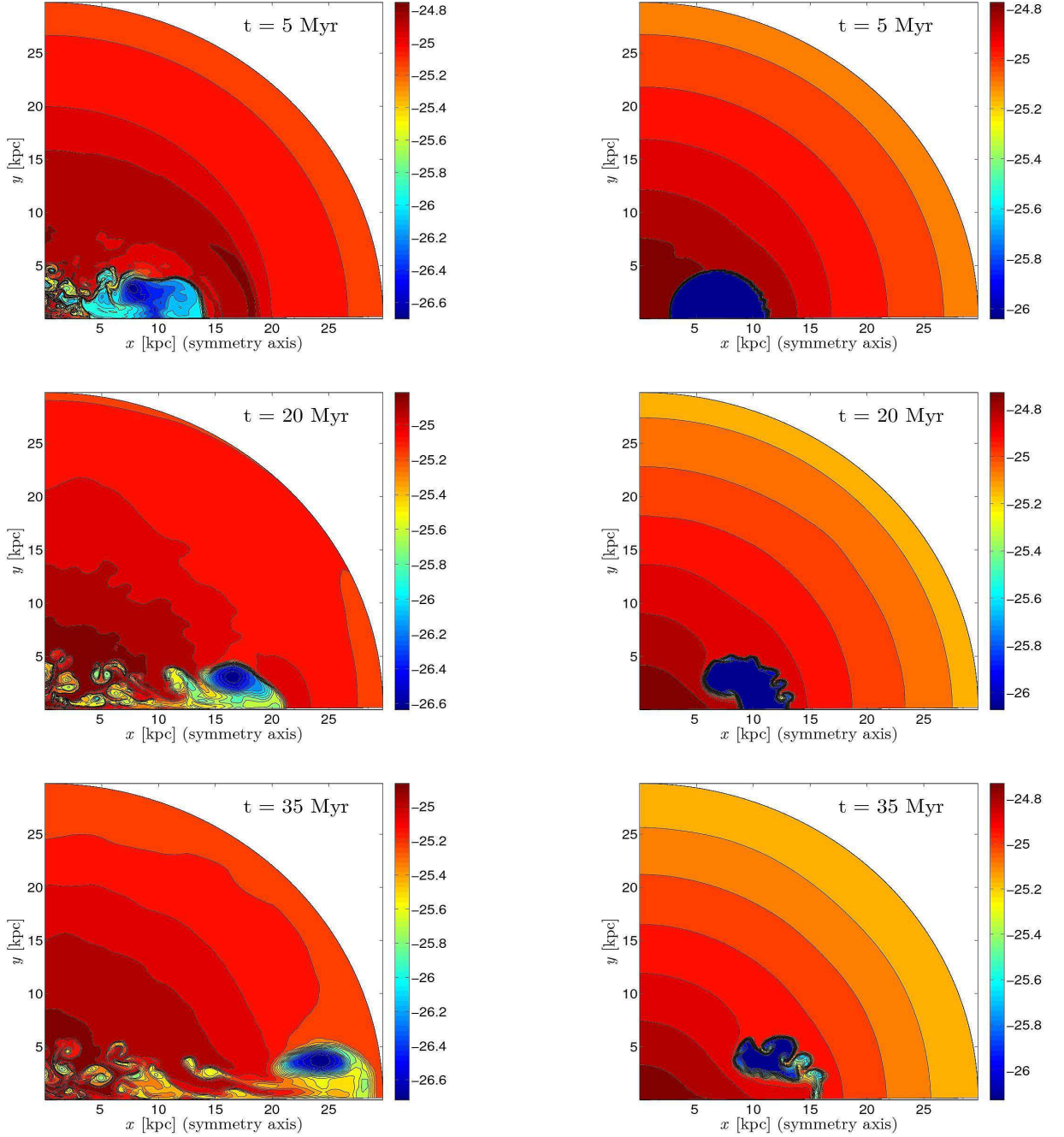

{\includegraphics[scale=0.29]{assafexj5f2a.eps2}} \hskip -1.5 cm
{\includegraphics[scale=0.29]{assafexb5f2b.eps2}} \\
{\includegraphics[scale=0.29]{assafexj20f2c.eps2}} \hskip -1.5 cm
{\includegraphics[scale=0.29]{assafexb20f2d.eps2}} \\
{\includegraphics[scale=0.29]{assafexj35f2e.eps2}} \hskip -1.5 cm
{\includegraphics[scale=0.29]{assafexb35f2f.eps2}}
\caption{The evolution of the jet-inflated bubble (left column) and the
artificial bubble (right-column), at three times as indicated.
In the evolution of the jet-inflated bubble the jet was active from
from $t=-10 \Myr$ until $t=0$, when the jet was completely shut-off.
The artificial bubble was introduces as a spherical bubble with a constant 
density at $t=0$.}
\label{fig:compare}
\end{figure}

In Figure \ref{fig:compare} we compare the evolution of the jet-inflated bubble
with that of an artificial bubble.
As is well documented in many non-magnetic simulations of artificial bubbles
(e.g., Churazov et al. 2001; Robinson et al. 2004; Reynolds et al. 2005;
Jones \& De Young 2005), as the bubble rises a vortex from below forms a flow
along the axis that penetrates the bubble from below. This flow leads to the 
destruction of the bubble. In the case of an artificial bubble this occurs 
over a short distance, and the bubble clearly loses its shape.

With the more realistic jet-inflated bubble the situation is very different:
\begin{enumerate}
\item RT instability modes are seen on the front of the artificial bubble.
The jet-inflated bubble does not suffer from such instabilities at early
times. This is a result of the outflow velocity of the dense shell in front
of the bubble. This interface is stable during the inflation phase (Soker et
al. 2002; Pizzolato \& Soker 2006), and the stability is maintained as long
as the low density bubbles does not support the dense ICM against gravity.
\item The jet-inflated bubble reaches a much larger distance in the cluster
before it starts losing its shape, as compared with the artificial bubble.
This is a result of the momentum of the bubble and the dense shell around it.
\item The jet material that lags behind the bubble, partially fills the
region along the symmetry axis. Our numerical grid forces the flow to be
exactly axisymmetric.  In a more realistic 3D flow, we expect this region to
be spread around the symmetry axis, such that in projection the a bubble will
still be observed as a more or less spherical bubble. In the case of an
artificial bubble this region is filled with ICM dense gas, and in
projection the bubble will appear as a torus.
{{{ In addition, in a future paper we intend to follow Reynolds et al. (2005) 
and increase the viscosity. We expect the higher viscosity to reduce the flow 
of dense matter along the symmetry axis. }}}
\item It seems that the vortex inside the jet-inflated bubble (Fig.
\ref{fig:jeti}) stabilizes the sides of the bubbles as it rises. We see no
sign of instabilities there, neither RT nor Kelvin-Helmholtz.
\item There is a low density-high temperature (high entropy) gas lagging
behind the bubble in a disrupted flow. This gas is mixed with the ICM and
increase its entropy. What we find here is a relatively efficient way to
heat the ICM.
\end{enumerate}

{{{{We would like to state that our results also support past conclusions 
that simulations of feedback in cooling flow clusters have to be more realistic
and incorporate, among other things, a more realistic jet (Vernaleo \& 
Reynolds, 2006). The inflation of the bubbles by the jet, and not by
predescribed numerical recipe, might be one of the missing ingredients needed 
to make these simulations more real.}}}}

\section{DISCUSSION AND SUMMARY}
\label{sec:discussion}

When the realistic jet-inflation process of bubbles is considered there
are two stabilizing processes.
(1) During the inflation phase the bubble is RT stable because the bubble-ICM
interface is decelerating and expanding (Soker et al. 2002; Pizzolato \& Soker 
2006).
(2) The outward momentum of the bubble and the dense shell around it implies
that the low density bubble does not need to support a dense gas above it
during the outward motion to large distances. This implies that the interface
is RT stable. The result of these processes is that the bubbles can rise to a
large distances from the cluster's center while still maintaining their 
general structure, without the need to invoke stabilizing magnetic field. 
Therefore, it is crucial that in the study of low-density bubbles in clusters 
of galaxies, the bubbles will be self-consistently inflated, rather than 
introduced artificially. It seems also that the vortex within the jet-inflated 
bubble suppresses instabilities on the sides of the rising bubble.

Our results show that the inflating jet, even if completely shut down, result
in a high entropy gas extending from the center to the bubble. This
high-entropy gas mixes with the ICM. This mixing is a relatively efficient
channel to heat the ICM. As seen at late times in Figure \ref{fig:compare},
the volume filled by the `fractal' high entropy gas is non-negligible.

The inflation of large, more or less spherical, bubbles close to the center
of the cluster (termed `fat' bubbles) requires the jet to be slow,
$v_j \sim 10^4 \km \s^{-1} \ll c$, and the mass loss rate to be relatively
large, $\sim 1-50 M_\odot \yr^{-1}$ (Sternberg et al. 2007; Sternberg \&
Soker 2008). In these papers we already mentioned AGN observations that
support such a high mass loss rate, and listed some previous theoretical
studies based on such an outflow. We here add that recent observations suggest
that in many clusters, cooling flow does exist as the radiative cooling is
not completely prevented (O'Dea et al. 2008). Namely, gas radiatively cools
to low temperatures. Revaz et al. (2008) conducted numerical simulations and
use them to suggest that cold blobs of gas participate in the feedback heating
in cooling flow clusters, as suggested by Pizzolato \& Soker (2005). Over all,
it seems that high mass loss rate in jets might occur quite often in cooling
flow clusters, as suggested by the moderate cooling flow model (Soker \&
Pizzolato 2005).
{{{{Another issue is the opening angle of the jet. We note that most 
observed jets are narrow. However, as discussed in Sternberg \& Soker (2007),
it is possible that much of the energy and mass reside a wider and slower 
outflow, i.e., the narrow jet is the center of a wider jet. Our prediction 
of a wide, or rapidly precessing, jet will have to be tested in the future.}}}}

Putting the results of this paper on a broader view, the usage of slow massive
jets, that are either wide of rapidly precessesing, can account for some basic
properties of cooling flow clusters:
(1) Recycling of gas that cool from the ICM and flows toward the center.
(2) Formation of fat bubbles close to the center.
(3) Allowing the bubble to rise to large distances (the results of this paper).
(4) Efficiently transfer energy form the central accreting black hole to the
ICM.

\acknowledgements
We thank John Blondin for his immense help with the numerical code. {{{{We 
thank Chris Reynolds (the referee) for helpful comments, and Edward Pope for 
his insightful remarks.}}}} This research was supported
by the Asher Fund for Space
Research at the Technion.

\end{document}